# Growth and electronic and magnetic structure of iron oxide films on Pt(111)

N. Spiridis[1], D. Wilgocka-Ślęzak[1], K. Freindl[1, 2], B. Figarska[1], T. Giela[2], E. Młyńczak[2], B. Strzelczyk[2], M. Zając[3], J. Korecki[1,2]

[1]Jerzy Haber Institute of Catalysis and Surface Chemistry PAS, Kraków, Poland

[2] Faculty of Physics and Applied Computer Science, AGH University of Science and Technology, Kraków, Poland

[3] European Synchrotron Radiation Facility, BP220, F-38043 Grenoble, France

## Abstract

Ultrathin (111)-oriented polar iron oxide films were grown on a Pt(111) single crystal either by the reactive deposition of iron or oxidation of metallic iron monolayers. These films were characterized using low energy electron diffraction, scanning tunneling microscopy and conversion electron Mossbauer spectroscopy. The reactive deposition of Fe led to the island growth of $Fe_3O_4$, in which the electronic and magnetic properties of the bulk material were modulated by superparamagnetic size effects for thicknesses below 2 nm, revealing specific surface and interface features. In contrast, the oxide films with FeO stoichiometry, which could be stabilized as thick as 4 nm under special preparation conditions, had electronic and magnetic properties that were very different from their bulk counterpart, wüstite. Unusual long range magnetic order appeared at room temperature for thicknesses between three and ten monolayers, the appearance of which requires severe structural modification from the rock-salt structure.

**PACS**: 68.47.Gh Oxide surfaces, 68.55.-a Thin film structure and morphology, 75.70.Ak Magnetic properties of monolayers and thin films, 81.15.-z Methods of deposition of films and coatings; film growth and epitaxy, 82.80.Ej X-ray, Mössbauer, and other γ-ray spectroscopic analysis methods

**Keywords**: iron-oxide films, Pt(111), FeO(111), Fe3O4(111), Pt(111), LEED, STM, CEMS

# 1. Introduction

Metal oxide surfaces and thin films are subjects of great interest because of their broad spectrum of technological applications in different areas [1]. Thin films of the transition metal oxides, such as iron oxides, are unique because in addition to their catalytic properties [2] they have magnetic applications [3]. Depending on their stoichiometry and structure, the catalytic and magnetic properties of these oxides can vary significantly. The character of the magnetic structure (paramagnetic, ferromagnetic or antiferromagnetic) is determined by the crystal structure and composition. As a result, the magnetic properties can be tuned in nanostructures and thin films by special preparation conditions and thickness [4]. The magnetic properties of thin films can differ greatly from those of the bulk [5]; this may have important consequences for possible spintronic applications.

Magnetite is one of the best candidates for such applications. Grown on different substrates and by various methods, magnetite thin films have been widely described in the literature (for recent literature references, consult Refs. [6, 7, 8]). Many of these studies have focused on magnetite films grown on MgO(001) surfaces by the reactive deposition of iron in the atmosphere of molecular oxygen, which predominantly leads to the stabilization of (001)-oriented $Fe_3O_4$ films. The complex electronic and magnetic properties of magnetite [9] also contribute to the complexity of the termination and reconstruction of the polar $Fe_3O_4$(001) surface [10], but special preparation recipes allow the surface structure and composition to be controlled [11]. Apparently, polarity compensation does not influence the epitaxial growth, and flat and continuous (001)-oriented magnetite films can be epitaxially grown without thickness limitation [12, 13]. This system, however, exhibits undesirable magnetic properties associated with structural defects (anti-phase domain boundaries), such as superparamagnetism [14, 7] and very high saturation field [15].

On the other hand, (111)-oriented magnetite films have been stabilized on simple metal (111) surfaces [8], in particular on Pt(111) (for a review, see Ref. [16]). Low energy electron diffraction (LEED) and scanning tunneling microscopy (STM) studies concerning iron oxides grown by post-deposition oxidation of metallic iron were presented by Ritter et al. [17]. These studies showed that, for the initial monolayers (one, two or three, depending on preparation temperature), an FeO(111)-like phase is stabilized, and the $Fe_3O_4$ (magnetite) phase is formed in the following layers. In contrast to the (001)-oriented films on MgO(001), these $Fe_3O_4$(111) films exhibit Stranski-Krastanov island growth and a variety of surface structures [18, 19]. It is likely that the dimension reduction produced by the formation of nano-sized islands promotes the stabilization of the polar, charge-uncompensated terminations [20]. The predominance of the magnetite phase in thicker layers was also verified in a series of experiments using conversion electron Mössbauer spectroscopy (CEMS) [21, 22], but the interpretation of the phase composition near the platinum substrate was ambiguous. It remains unknown whether the interface FeO layer is preserved or transforms to magnetite in thicker films.

The studies of iron oxide films with the wüstite stoichiometry are of general importance. Wüstite is a non-stoichiometric $Fe_{1-x}O$ iron oxide that crystallizes in the rock salt crystal structure. As a basic oxide component of the interior of the Earth, wüstite has been subjected to numerous high pressure measurements that reveal the remarkable sensitivity of its structural, electrical and magnetic properties to stoichiometry and interatomic distances [23, 24]. Because of this sensitivity, the epitaxial stresses in the FeO film prepared on Pt(111), produced by the considerable mismatch of the atomic spacing between the platinum (2.78 Å) and wüstite (3.04 Å) (111)-planes, can be expected to have a significant effect on the film properties. Additionally, the stabilization of (111)-oriented FeO films beyond the limit of 2 ML is important to understand the mechanism of polarity compensation in (111) oxide films on a metal substrate [25].

The aim of this paper is to systematize and directly compare the growth and properties of iron oxide thin films that are prepared on the Pt(111) substrate in two ways: by post-deposition oxidation of metallic iron and by reactive deposition of Fe in an oxygen atmosphere. Unambiguous phase identification is essential for such a comparison. Because of their surface-limited sensitivity, the STM and LEED methods are not sufficient for this purpose. Their supplementation by *in situ* Mössbauer spectroscopy seems to be the most reasonable choice for studying iron containing systems. CEMS, with its submonolayer sensitivity, is a powerful method that not only enables phase analysis but also gives local in-depth information on the electronic and magnetic (also antiferromagnetic) state. This is especially useful for studying subtle ultrathin film magnetism, taking into account such issues as the size effect, which is strongly dependent on the film thickness [26], and the complex interplay between the structure, strain and magnetic order [5].

In the present paper, we show that both FeO (ferrous) and $Fe_3O_4$ (magnetite) thin films can be stabilized on Pt(111); however, they exhibit different growth modes and relations to the properties of their bulk counterparts. This observation is relevant to understanding the conditions under which ultrathin oxide films behave like the bulk material [27] and where the borderline between 2D and 3D behavior occurs.

## 2. Experimental details

Iron oxide films were grown in a multipurpose UHV apparatus with a base pressure of $1 \times 10^{-10}$ mbar. The apparatus is equipped with a molecular beam epitaxy (MBE) system for deposition of $^{56}$Fe and $^{57}$Fe isotopes and standard surface characterization methods: LEED and AES. In separate chambers of the same apparatus, scanning tunneling microscopy (Burleigh Instruments) and CEMS were used for *in situ* sample analysis. The CEMS measurements were performed with a 54° angle between incident gamma radiation and the sample surface normal. The CEMS spectra were numerically fitted using the Voigt lines (convolution of the Lorentzian and Gaussian), a technique that allows straightforward and consistent implementation of the hyperfine parameter distribution, which is inherent in low-dimensional systems.

The Pt(111) substrate was cleaned by the standard procedure of Ar$^+$ bombardment cycles, oxygen atmosphere annealing and flashing until the sharp (1x1)-Pt(111) LEED pattern was observed, and no impurities were visible in the AES signal.

Two different recipes were followed to grow iron oxide layers over a wide range of thicknesses: reactive deposition in an oxygen atmosphere and post-deposition oxidation of metallic Fe monolayers. In all cases, iron enriched to contain 95% of the $^{57}$Fe isotope was used to facilitate the Mössbauer measurements. The deposition of iron was controlled by a quartz thickness monitor with the accuracy of approximately 0.2 ML. For estimation of the oxide layer thickness, the following equivalences were assumed: 0.1 nm of deposited metallic Fe provides 0.21 nm of Fe$_3$O$_4$ or 0.18 nm of FeO. Both oxides (wüstite and magnetite) are well matched to the Pt(111) substrate in the (111) orientation, and the atomic layer stack for both oxide structures along the [111] direction is shown in Fig. 1.

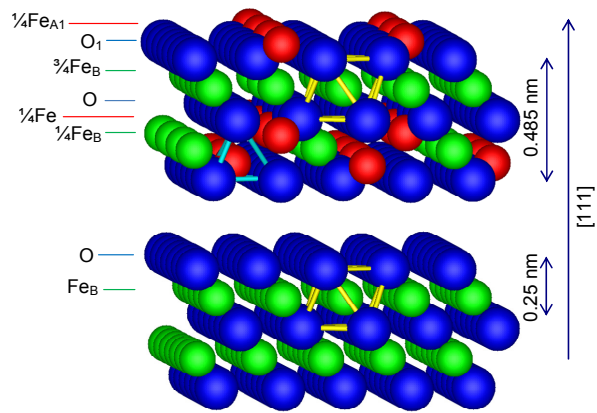

Fig. 1. The stacking of the (111)-atomic layers in spinel (magnetite, top) and rock-salt (wüstite, bottom) structures

## 2. Iron oxides by reactive deposition - formation of Fe$_3$O$_4$

### a. LEED and STM

Similar preparation conditions to those previously used for iron oxide films grown on a MgO(001) substrate, resulting in the epitaxial Fe$_3$O$_4$ phase [28], were now applied to the Pt(111) substrate. The films were prepared by deposition of iron in an oxygen atmosphere under a partial pressure of $8 \cdot 10^{-6}$ mbar and a substrate temperature of 550 K. The rate of iron evaporation was approximately 0.2 nm/min. Several samples, with nominal iron contents between 0.17 nm and 5 nm, were prepared. After deposition, the samples were UHV annealed at 800 K for 10 minutes.

The thinnest sample was reactively deposited using 0.17 nm of metallic Fe, which nominally corresponds to 1.2 ML FeO or 0.7 ML of $Fe_3O_4$, assuming their (111) orientation. The LEED pattern of this sample (Fig. 2a) exhibits six-fold symmetry with characteristic moiré-type satellites, which are produced by the lattice mismatch between the wüstite monolayer (a = 3.04Å) and the platinum substrate (a = 2.77 Å). The pattern is typical for a 1 ML FeO film being a polar bilayer formed by the iron monolayer neighboring with the Pt substrate and the oxygen monolayer that terminates the surface [16]. The moiré pattern is clearly visible in the STM image (Fig. 2d), which also reveals the formation of a second oxide layer in the form of irregular polygonal islands, in accordance with the nominal coverage. The islands cover approximately 30% of the area of the first layer, and their apparent height above the first oxide layer is 0.28 nm. The second atomic layer does not display the moiré pattern and at this stage, it is unclear whether the second layer is of the FeO or $Fe_3O_4$ type.

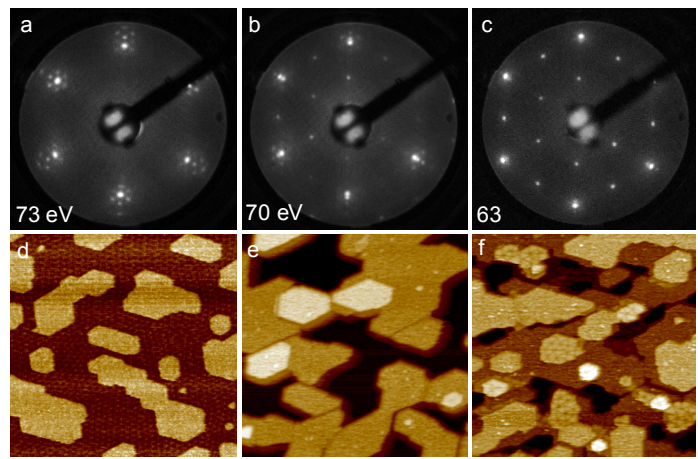

Fig. 2. LEED patterns and STM images ($100x100$ nm$^2$) for iron oxide films grown on Pt(111) by reactive deposition of iron: a and d – 0.17 nm, b and e – 0.5 nm, c and f – 0.8 nm.

With the increasing film thickness, qualitative changes in the LEED pattern were observed, indicating the formation of the $Fe_3O_4$(111) phase. For 0.5 nm of Fe (nominally ~1 nm of $Fe_3O_4$), the satellites around the (1x1) spots became weaker, and the (2x2) superstructure spots appeared (Fig. 2d). Such a doubling of the surface unit cell appears for the magnetite $Fe_3O_4$(111) surface with so called regular termination [29]. The corresponding STM images (Fig. 2e) show a continuous first oxide layer on which large islands have begun to coalesce, and on which an additional layer is visible. Most of the islands have an apparent height of approximately 1.5 nm, as measured from the STM image section. This distance, setting the periodicity along the [111] direction, corresponds to three (111)-physical monolayers (PMLs) of $Fe_3O_4$(111), where a single (111)-PML is understood to consist of a stack of two oxygen layers, two tetrahedral-Fe layers and two octahedral-Fe geometrical (atomic) layers with a total thickness of 0.486 nm (see Fig. 1). The topmost layer is formed by islands that protrude 0.5 nm above the 1.5 nm level. The height of the islands is quantized; it is an integer multiple of the PML thickness. The total coverage, as estimated from the STM image analysis, exhibits good correspondence to the amount of deposited iron (0.5 nm) within the assumed phase composition. The first continuous FeO-like layer consumes 0.14 nm of iron, and the total volume of the $Fe_3O_4(111)$ islands yields an average coverage of 1.6 PML, consuming the remaining 0.36 nm of Fe. As a whole, typical Stranski–Krastanov growth was observed, in which further growth proceeds via three-dimensional islands on the continuous FeO-like monolayer.

For thicker layers (prepared using 0.8 nm, 1.5 nm and 5.0 nm of $^{57}Fe$), the growth continues in the same way, as exemplified in Fig. 2f for the sample with an iron content of 0.8 nm [nominally 3.4 PML of $Fe_3O_4(111)$]. The LEED pattern is dominated by spots that are typical of the $Fe_3O_4(111)$ surface. Weakening satellites produced by FeO are visible only for the 0.8 nm Fe sample. With increasing thickness, the STM images show the increasing height amplitude of the magnetite islands. The atomic resolution images (not shown) display 0.6 nm atomic periodicity on most island surfaces, which is characteristic of (1/4 ML)-tetrahedral termination and is also a superstructure characteristic of oxygen deficient areas [29].

b. CEMS results for $Fe_3O_4$ films

All films were analyzed in situ by means of conversion electron Mössbauer spectroscopy to conclusively identify their oxide phase and probe their magnetic properties. The results of the RT Mössbauer measurements and their analyses are collected in Fig. 3. It seems reasonable to start the analysis with the spectrum of the thickest film (5 nm of $^{57}Fe$, nominally producing 10.5 nm of $Fe_3O_4$), which has a strong character of bulk magnetite. The Mössbauer spectrum of bulk magnetite at room temperature is characterized by two sextets. One, denoted "A", has a hyperfine magnetic field of $B_{hf} = 48.8$ T and an isomer shift of IS = 0.27 mm/s relative to α-Fe; it corresponds to the $Fe^{3+}_A$ ions at the tetrahedral A-sites. The second one, denoted "B" has $B_{hf} = 45.7(2)$ T and IS = 0.65 mm/s; it is the signal with $Fe^{2.5+}_B$ character from the cations at the octahedral B sites. $Fe^{+2}_B$ and $Fe^{3+}_B$ are indistinguishable because of fast electron transfer (electron hopping) [30]. Very similar components with minor modifications of their hyperfine parameters as a result of finite size and surface effects [31] constitute 95% of the spectral intensity. The remaining 5% (corresponding to no more than 1 ML) is in a single line with IS=0.24 mm/s, similar to that found previously by Schedin *et al.* [21]. This line has been interpreted as coming from an FeO monolayer at the $Fe_3O_4$/Pt interface or, more recently, as from iron atoms dissolved in the Pt substrate [22]. The later interpretation considers the possible accumulation of Fe dissolved in the Pt substrate after many preparation cycles. In the case of the *in situ* measurements in this study, such an effect definitely excluded by CEMS measurements performed on the Pt(111) substrate in a cleaned state, which never showed any resonance signal.

As the film thickness decreases, the spectra show pronounced deviation from those of the bulk. This change is manifested in two types of effects: (i) the magnetically split components become broader, less resolved and characterized by smaller hyperfine magnetic fields and (ii) the relative intensity of the central single line increases. The first type of effect is the size effect produced by the finite film thickness and the island structure. The distribution of the hyperfine parameters is caused by the broken translation symmetry normal to the film (enhanced surface contribution) [31] and in the film plane, which results in differentiated local coordination of the iron atoms exposed at the surface and at island boundaries. The reduced thickness is also reflected in the reduced Curie temperature [32], which in turn produces smaller value of the local magnetization (measured by the hyperfine magnetic field). Also superparamagnetism plays an essential role in blurring the Mössbauer spectra, as it was demonstrated recently for $Fe_3O_4(001)$ films thinner than 5 nm [7]. The superparamagnetic relaxation is blocked by lowering the temperature, and the CEMS spectra at cryogenic temperatures reveal typical features of magnetite (an example for a 2 PML film is shown in the inset of Fig. 3).

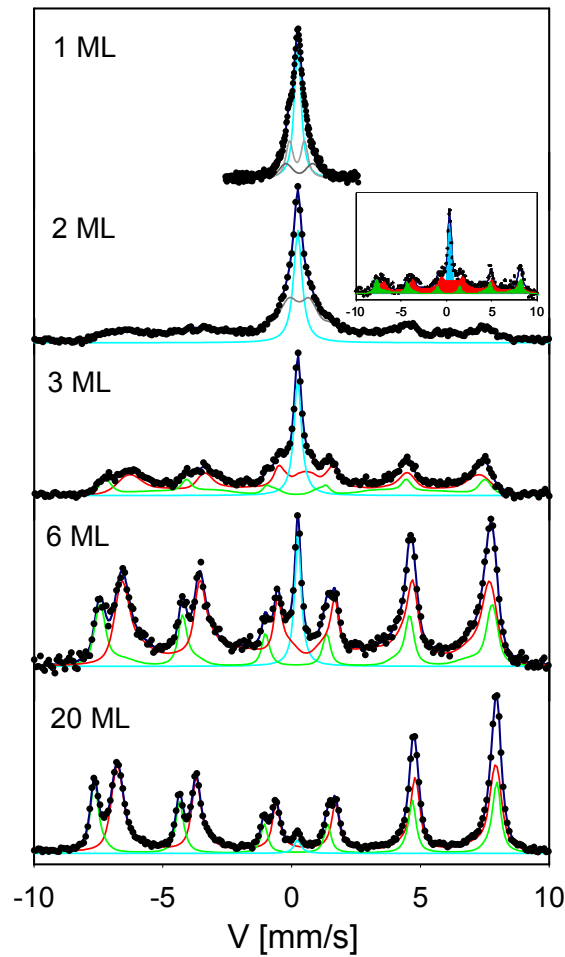

Fig. 3. Evolution of room temperature CEMS spectra with increasing thickness for iron oxide films grown on Pt(111) by reactive deposition of iron. The inset of the 2 ML spectrum shows the effect of low temperature (108 K). The solid lines show the result of the best fit and the deconvolution into spectral components. Red and green lines correspond to components unambiguously identified as coming from magnetite, and blue and magenta correspond to components identified as coming interfacial oxide-Pt(111) phases.

The second effect, namely the intensity increase of the central single line, supports the interpretation (base also on the LEED and STM data) that the single line has its origin at the interface, where one oxide monolayer of a specific stoichiometry (FeO) is formed, in analogy with previous studies of magnetite films prepared by the oxidation of metallic iron [18]. The properties of the FeO phase will be discussed in more detail in the next section.

## 3. Iron oxides obtained by oxidation of metallic-Fe monolayers

Usually, iron oxide thin films on Pt(111) are grown by post-preparation oxidation of consecutive metallic Fe monolayers in an $O_2$ atmosphere at an elevated temperature [ 17, 19, 21, 33, 34, 35, 36]. The typical growth conditions for these films consist of oxygen pressures between $10^{-7}$ mbar and $10^{-6}$ mbar and temperatures between 830 K and 1000 K. Films prepared in this way have a complex structure that depends on their thickness and preparation condition (oxygen pressure and substrate temperature) [16], as well as their post-preparation treatment [37, 38]. A consensus exists that, at the initial growth stage, a wüstite-like FeO(111) mono- and bi-layer are stabilized; beyond this, the layers have the $Fe_3O_4$ stoichiometry.

Ritter et al. [17] reported that the completion of the second and third FeO layer, as well as the characteristics of $Fe_3O_4(111)$ island growth that starts between 2 and 3 ML of FeO coverage, critically depends on the film oxidation temperature. Following this observation, we used lower temperature and oxygen partial pressure than the values specified in the typical parameters during monolayer oxidation to further suppress the formation of magnetite. Metallic iron monolayers (1 ML corresponded to 0.14 nm, as estimated by the quartz monitor) were deposited on the Pt(111) substrate at room temperature. Only the first iron monolayer was oxidized at $1 \times 10^{-6}$ mbar of $O_2$ and a substrate temperature of 850 K for 2 minutes. Starting from the second layer, different oxidation conditions were employed: a temperature of 570 K and an oxygen exposure of 10 L at $5 \times 10^{-8}$ mbar. Finally, the sample was annealed under UHV for 10 minutes at 900 K to allow temperature driven rearrangement at the atomic scale.

### a. LEED and STM.

The initial growth of FeO is essentially the same as that reported previously [17], with typical LEED patterns for the first two FeO monolayers (not shown), revealing a moiré superstructure produced by the mismatch between the stretched oxide monolayer and the Pt substrate. With increasing film thickness, the LEED pattern gradually simplified by disappearance of the substrate features, transforming to a simpler six-spot symmetric satellite pattern around the main FeO(111) spots, as shown in Figs 4a-c. The superstructure can eventually be described as (7x7) with respect to the (1x1)FeO(111) pattern, with an in-plane lattice constant of 3.14±0.01 Å, which corresponds to a superstructure period of approximately 22 Å. Then, at a thickness between 7 ML and 10 ML, a second structural domain appears, as manifested by the second set of spots rotated by 30°. The character of these new features changes slightly from preparation to preparation (this observation is derived from five independent preparation runs) in an uncontrolled way. Sometimes, the second set of spots displays the superstructure as described above; however, its maximum intensity occurs at slightly different energy and focus conditions. Sometimes the structural domains also appear different at different macroscopic sample positions.

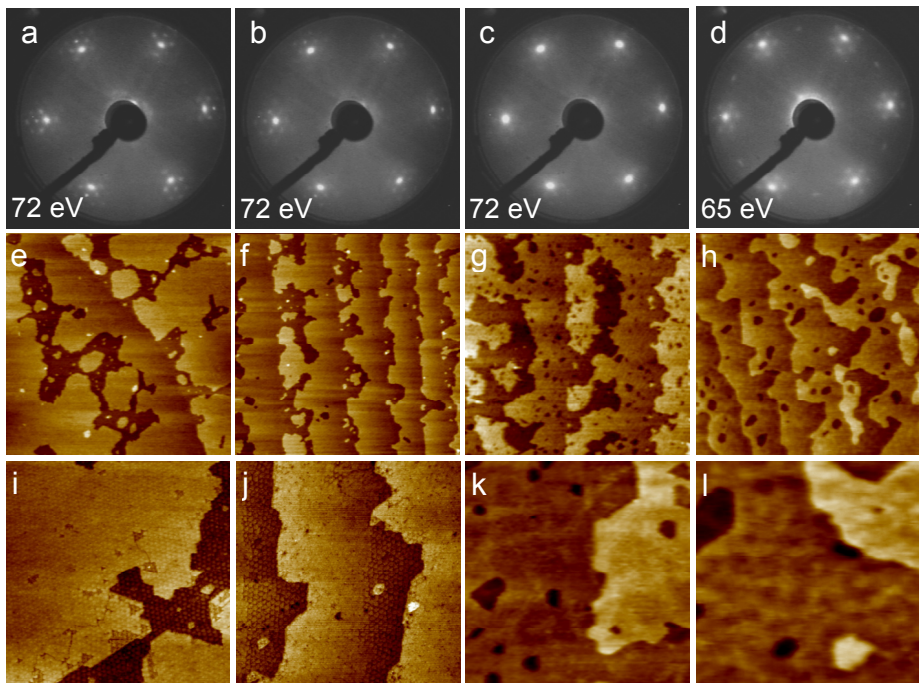

Fig. 4. LEED patterns and STM images (400x400 nm$^2$, middle and 100x100 nm$^2$, bottom) for iron oxide films grown on Pt(111) by oxidation of iron monolayers. The columns corresponds to the nominal oxide thickness of 3 ML, 4 ML, 5 ML and 7 ML, from left to right, respectively.

Additional information comes from the STM images. Initially, the films grew flat, and we observed the coincidence structures that are typical for the first and second monolayers [17]. However, in contrast to the previous studies, the flat growth continued beyond the second monolayer, as clearly shown in the STM images in Fig. 4. The size of the atomically flat terraces corresponds to that of the Pt substrate, and only the two topmost layers are typically exposed. The STM-measured step height was very similar for all thicknesses, close to the value of 2.5 Å expected for the (111)-oriented Fe-O bilayers in the bulk rocksalt-FeO structure. For the third and fourth monolayers (Fig. 4i,j), a regular moiré pattern is visible with a periodicity of 23±1 Å, in agreement with the LEED patterns. For thicker films, the acquisition of the STM images became more laborious, and the atomic scale periodic corrugations were replaced by irregular ripples with a ~0.5 Å amplitude (Fig. 4k,). Both the amplitude and the pattern of these ripples were strongly dependent on the thickness. The ripples seem to have topographic (not electronic) character, and they are an effect of buckling due to epitaxial stress or polarity compensation. For the thickest films (approaching 20 ML), the STM scans became very unstable and produced fuzzy images.

The surface symmetry of our iron oxide films, as observed with LEED and STM, indicates that the growth of oxygen-terminated FeO(111) continues over 10 ML for specific preparation conditions. The stoichiometry and the surface structure of these films are very different from those of the $Fe_3O_4$(111) films on Pt(111) described in the previous section. For the magnetite films, we determined the AES signal ratio of the 510 eV oxygen and 651 eV iron lines to be R = 3.8(1), while the thick FeO films had an R value of 2.9(1), proportionally to the expected oxygen-iron stoichiometric ratio. However, despite the wüstite stoichiometry, the electronic and magnetic properties determined by CEMS measurements of the films are different from those expected for FeO films of the NaCl structure, as described in the next subsection.

b. CEMS results for FeO films

The Mössbauer spectra of our FeO films are specified in reference to the wüstite spectra. Wüstite adopts the rock salt structure above its Néel temperature ($T_N \approx 198$ K). However, it is well known that FeO is non-stoichiometric, accommodating a cation deficiency by the formation of octahedral iron vacancies and a small number of tetrahedral iron(III) interstitials. These defects tend to aggregate and form tetrahedral units, which were identified by neutron diffraction and Mössbauer spectroscopy [39, 40]. The bulk magnetic properties of wüstite $Fe_yO$ are complex; it is an antiferromagnet with an exact Néel temperature that depends on $y$ [41]. Below the magnetic ordering temperature, $Fe_yO$ undergoes a rhombohedral distortion, and the iron spins align along the [111] direction of the unit cell, forming antiferromagnetically coupled alternate (111) iron ferromagnetic sheets [42]. Considering the cubic structure, the Mössbauer spectrum under ambient conditions should contain one singlet corresponding to $Fe^{2+}$ in the octahedral site; however, due to the non-stoichiometry of the material, the room temperature spectra show several singlets and doublets corresponding to undistorted octahedral $Fe^{2+}$ sites, octahedral $Fe^{2+}$ sites associated with vacancies and complex defect clustering, and also $Fe^{3+}$ in octahedral and tetrahedral positions [24]. The doublet that dominates the spectrum is a fingerprint of the wüstite Mössbauer spectrum, with a relatively high isomer shift value of IS $\approx$ 0.9 mm/s and a distinct quadrupole splitting of QS $\approx$ 0.6 mm/s.

The CEMS spectra of our FeO films, whose evolution with increasing thickness is shown in Fig. 5, are indicative of non-wüstite phase. The most characteristic features of the spectra set are the following: (i) an isomer shift in the range of 0.3 mm/s of the dominant spectral components, (ii) small or negligible quadrupole splitting and (iii) a magnetic order that appears in a certain thickness range.

The solid lines in Fig. 5 are the results of numerical fits of the CEMS spectra, including their decomposition into spectral components. The spectra were fitted with consideration of consistency in the evolution of spectral components with increasing thickness. The CEMS spectrum for one (not shown) and two FeO monolayers are similar, presenting a slightly asymmetric line that can be best fitted with a single line centered at IS = 0.35(1) mm/s and broader features that can be described as one or two quadrupole doublets with QS ranging between 0.5 mm/s and 0.9 mm/s and IS between 0.3 mm/s and 0.4 mm/s. A similar spectrum was also observed for the monolayer oxide film prepared by the reactive deposition of iron (Fig. 3 in the previous section). Considering the high structural quality of the monolayer iron oxide films, the diversification of the spectral component can be explained by the different positions of the iron atoms in the modulated FeO coincidence structures. The analysis of this issue presents a topic for further study.

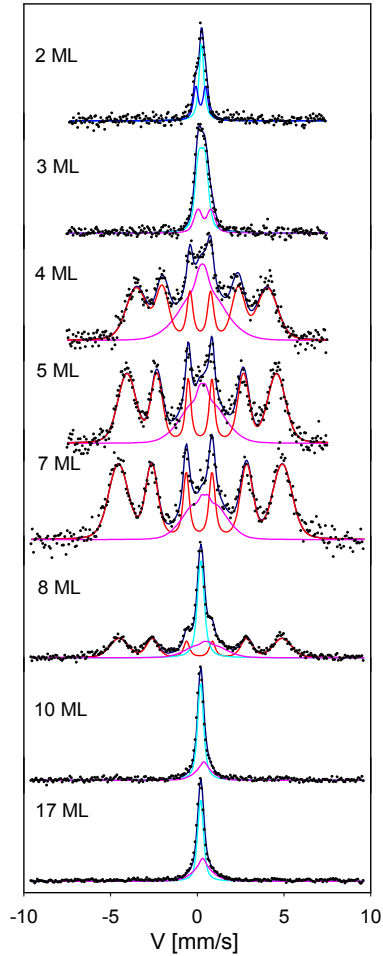

Fig. 5. Room temperature CEMS spectra for iron oxide films grown on Pt(111) by oxidation of iron monolayers as a function of increasing thickness. The solid lines show the result of the best fit and the deconvolution into spectral components: blue and red are the low isomer shift (LIS) components in the nonmagnetic and magnetic state, respectively, and purple is the high isomer shift (HIS) component.

Starting from the 3 ML thickness, the spectra reveal a long range magnetic order that is manifested in the 3 ML spectrum as a broadening of the central features and for the thicker films as a distinct six line magnetic pattern, which is superimposed with a Λ-shaped component, indicating a broad distribution of the hyperfine magnetic field. The six line pattern has an isomer shift close to that of the central single line (IS ≈ 0.3 mm/s), while the centre of gravity of the Λ-shaped component is shifted to a more positive velocity (IS in the range of 0.5 mm/s to 0.6 mm/s). Depending on the preparation run, the distinct magnetic features of the spectra were observed in the thickness range from 3 ML to 10 ML, and the different spectral components appeared slightly different. In particular, the single line that reappears in the presented data set for the 8 ML samples sometimes co-exists with the magnetic component in a wider thickness range. For the thickest films, 10 ML and above, the distinct magnetic component disappears, and the spectrum is dominated by the central single line. However, signs of the magnetic order remain in the form of a broader Λ-shaped satellite with a higher isomer shift.

The results of the numerical analysis of the FeO spectra over the entire range of thicknesses are summarized in Fig. 6. To describe the evolution of the samples with increasing thickness, we take the isomer shift value as the fingerprint of the chemical state of the Fe ions. In such an interpretation, the samples contain two types of iron, those with lower IS (LIS) and those with higher IS (HIS). The LIS component dominates, with an LIS to HIS ratio of 70 to 30 ($\pm$10). The predominant LIS component has a hyperfine pattern that is unusual for the high spin $Fe^{2+}$ ions expected in a bulk-like FeO structure, indicating the presence of another state (e.g., $Fe^{3+}$). While the LIS value is relatively stable across the entire thickness range, the evolution of the spectral magnetic features indicates the existence of sharp phase transitions. The onset of a long range magnetic order between 3 and 4 monolayers may be associated with a size effect that consists in an increase of the magnetic phase transition temperature with increasing thickness [26]. However, the origin and the character of the room temperature long range magnetic order (absent in bulk FeO) as well as its disappearance above 10 ML are intriguing. The hyperfine magnetic field ($B_{hf}$) of the dominating HIS component in the magnetically ordered state amounts to approximately 30 T, which significantly differs from the value that occurs for $Fe^{3+}$ high spin states (approximately 50 T). Assuming that only the contact term contributes to $B_{hf}$, the observed value would correspond to a spin magnetic moment of approximately 3 $\mu_B$ because the ground state estimate of $B_{hf}$ produced by 3d-polarization effects provides approximately 11 Tesla per unpaired 3d-spin [43]. From the CEMS data alone, the type of the magnetic order cannot be determined. However, the magnetooptic Kerr effect measurements of a 5 ML sample, in which a magnetic order was detected by CEMS, revealed a rectangular magnetic hysteresis loop, which is typical for ferromagnetism.

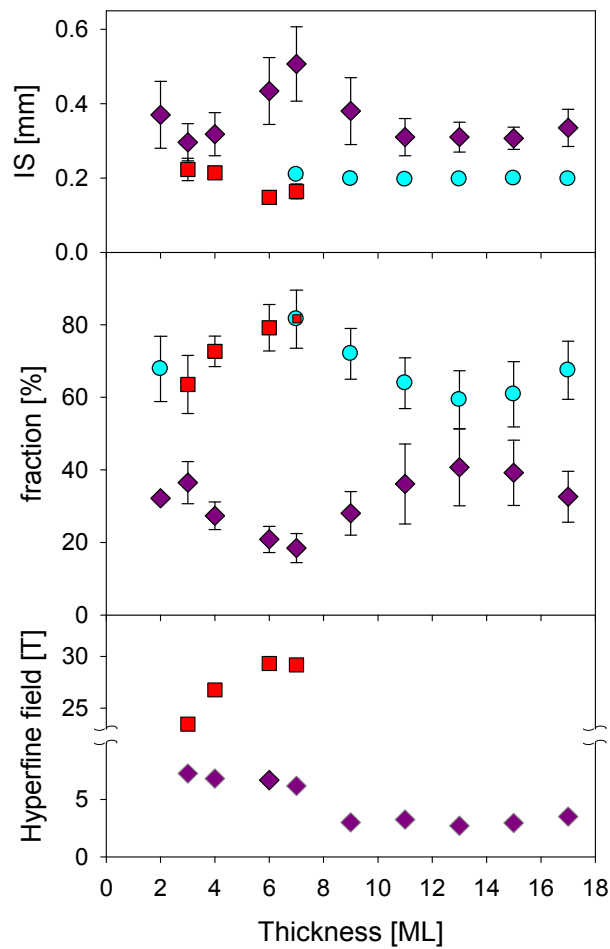

Fig. 6. Results of CEMS analysis for iron oxide films grown on Pt(111) by oxidation of iron monolayers: (top) Isomer shift (relative to α-Fe), (middle) fraction of spectral component and (bottom) hyperfine magnetic field as a function of the increasing thickness. Squares and circles are used for the low isomer shift (LIS) component in magnetic and non-magnetic state, respectively; diamonds are used for the high isomer shift (HIS) component.

The observation of room temperature ferromagnetism (unusual for FeO), together with the atypical hyperfine pattern, signifies electronic properties that are very different from those of bulk FeO. The magnetic moment estimated from the $B_{hf}$ value could suggest the high spin 2+ state of iron; however, the low value of IS is inconsistent with the well established systematics for the iron ionic compound [43]. The unusual IS can be explained by the covalency effect, which is understood as the contribution of 4s electrons to iron-oxygen bonding [44]. Such an effect would require a decrease of the Fe-O bond distance, which was directly confirmed for the first FeO monolayer [45] and may be well expected for the thicker films, especially if the stabilization of relatively thick polar films occurs via structural modification of the bulk phase. Such a situation was discussed theoretically for the MgO(111) case by Goniakowski et al. [46], who showed that a graphite-like structure could provide an alternative to the expected rock-salt structure. We also consider the existence of an alternative structure for FeO, moreover that the rock-salt phase in bulk must be stabilized by deviation from stoichiometry and that bulk FeO shows structural, electronic and magnetic phase transformations at high pressure [23, 47, 48, 49].

As stated above, the CEMS spectra indicated some inhomogeneity in the sample. The Λ-shaped HIS component with a broad distribution of $B_{hf}$ also shows a magnetic transition, but it is less sharp than the LIS transition. Moreover, numerical analysis proved that the magnetic order associated with the HIS component does not collapse entirely as the film increases in thickness. Rather, a small $B_{hf}$ of approximately 3 Tesla is characteristic of the entire thickness range between 10 ML and 17 ML. This component could be interpreted to indicate the presence of a phase with a different (e.g., antiferromagnetic) order, whose occurrence requires the atomic volume to change with the evolution of an epitaxial stress. Similar volume and structure dependence of the magnetic moment and hyperfine magnetic field has been observed in bcc- and fcc-iron [50]. Stress relaxation may also be responsible for the phase separation, as directly observed in LEED and CEMS and suggested by the ripple pattern observed in STM.

## 4. Conclusions

Preparation of iron oxide films on Pt(111) by different methods, namely, the reactive deposition of iron and the oxidation of metallic iron monolayers, produced magnetite ($Fe_3O_4$) and FeO-like phases, respectively. Over a broad thickness range (up to 10 nm), the $Fe_3O_4$ island films present many features of the bulk, with the exception of one interfacial monolayer at the Pt(111) substrate, which is nonmagnetic at room temperature and has FeO stoichiometry. The electronic and magnetic properties of the magnetite bulk phase are modulated only by superparamagnetic size effects for thicknesses below 2 nm, and they are influenced by some degree of disorder characteristic for low dimensional systems.

In turn, the FeO films, which could be stabilized in a flat and continuous form as thick as 4 nm (17 Fe-O bilayers) with careful optimization of the oxidation conditions, had very different electronic and magnetic properties from those of the bulk FeO with the rock-salt structure. The hyperfine interaction parameters were derived from the measured CEMS spectra, which usually provide good fingerprints of the oxide phase. These parameters cannot be associated with any compound existing in bulk. In particular, the isomer shift indicates a high degree of covalency in the Fe-O bonds. However, the most remarkable feature is the long range magnetic order (presumably ferromagnetic) observed in the thickness range of a few monolayers. Certainly, epitaxial strain related to the film-substrate lattice misfit and the polarity of the (111)-oriented films must play some role; however, taking into account the strength of the observed effects, we do not exclude the explanation that the resulting distortions might have induced a transition to a different structural phase.

## Acknowledgements

This work was supported in part by the Polish Ministry of Science and Higher Education and its grants for Scientific Research and by the Team and MPD Programs of the Foundation for Polish Science co-financed from the EU European Regional Development Fund.